\documentclass[%
 reprint,
 amsmath,amssymb,
 aps,
prb,
]{revtex4-1}

\usepackage{graphicx}
\usepackage{dcolumn}
\usepackage{bm}
\usepackage{multirow}
\usepackage{blindtext}
\usepackage{amsmath}
\usepackage{natbib}
\usepackage{lipsum}

\begin{document}



\title{\textit{Ab-Initio} Molecular Dynamics Acceleration Scheme with an Adaptive Machine Learning Framework}


\author{Venkatesh Botu}
\author{Rampi Ramprasad}%
\email{rampi@uconn.edu}
\affiliation{%
 University of Connecticut, Storrs, CT 06269
}
\date{\today}

\begin{abstract}
Quantum mechanics based \textit{ab-initio} molecular dynamics (MD) simulation schemes offer an accurate and direct means to monitor the time-evolution of materials. Nevertheless, the expensive and repetitive energy and force computations required in such simulations lead to significant bottlenecks. Here, we lay the foundations for such an accelerated \textit{ab-initio} MD approach integrated with a machine learning framework. The proposed algorithm learns from previously visited configurations in a continuous and adaptive manner on-the-fly, and predicts (with chemical accuracy) the energies and atomic forces of a new configuration at a minuscule fraction of the time taken by conventional \textit{ab-initio} methods. Key elements of this new accelerated \textit{ab-initio} MD paradigm include representations of atomic configurations by numerical fingerprints, the learning algorithm, a decision engine that guides the choice of the prediction scheme, and requisite amount of \textit{ab-initio} data. The performance of each aspect of the proposed scheme is critically evaluated for Al in several different chemical environments. This work can readily be extended to address non-elemental compounds, and has enormous implications beyond \textit{ab-initio} MD acceleration. It can also lead to accelerated structure and property prediction schemes, and accurate force-fields.
\end{abstract}

\maketitle


\section{\label{sec:level1}Introduction}

Computation-driven rational materials design efforts are rising in popularity and importance \cite{Ceder_1, Sharma_1}. This trend is being fueled by systematic improvements in capabilities to compute materials properties accurately and practically. Parameter-free (or \textit{ab-initio}) quantum mechanics (QM) based schemes such as density functional theory (DFT) are central to this unfolding development \cite{Neugebauer_1, Hautier_2, Becke_1}. While powerful, versatile, and efficient, \textit{ab-initio} methods are still too time-intensive to adequately handle several important classes of problems. For instance, the explicit dynamical evolution of materials and processes with timescales larger than a nanosecond are still beyond the reaches of DFT computations. 

The most direct way to handle and monitor the time-evolution of matter is by the molecular dynamics (MD) method \cite{Petrenko_1}. In \textit{ab-initio} MD, the ingredients necessary to perform MD, namely, the total potential energies and atomic forces are obtained using QM, but the evolution of the atoms (i.e., determination of the next new configuration, based on the current configuration, velocities and forces) is performed classically. The repetitive and expensive QM energy and force computations, and the necessity for small time-s.png (of the order of femtoseconds), lead to the primary bottlenecks of \textit{ab-initio} MD. Creative schemes to accelerate MD simulations so that longer timescales can be accessed have indeed been developed in the past\cite{Torrens_1,Henkelman_1,Chatterjee_1,Laio_1,Laio_2,Sorensen_1,Voter_1,  Voter_3, Voter_4, Voter_2, Hamelberg_1, Elliott_1}. These include the use of parameterized force-fields (rather than QM) to evaluate the energies and forces rapidly \cite{Torrens_1}, and/or “speeding the clock” using Monte Carlo methods \cite{Henkelman_1,Chatterjee_1}, meta-dynamics \cite{Laio_1, Laio_2}, temperature acclerated dynamics\cite{Sorensen_1,Voter_1,Voter_3,Voter_4} and hyperdynamics \cite{Voter_1, Voter_2, Voter_3, Voter_4}. These attempts though are not entirely satisfactory. Force-fields are not transferrable to situations that were not originally used in the parameterization, and altering the clock requires some prior knowledge of the critical features encountered during the evolution process (and involve artificial constraints and some loss of vital dynamical information).

\begin{figure*}
\includegraphics[width=7in]{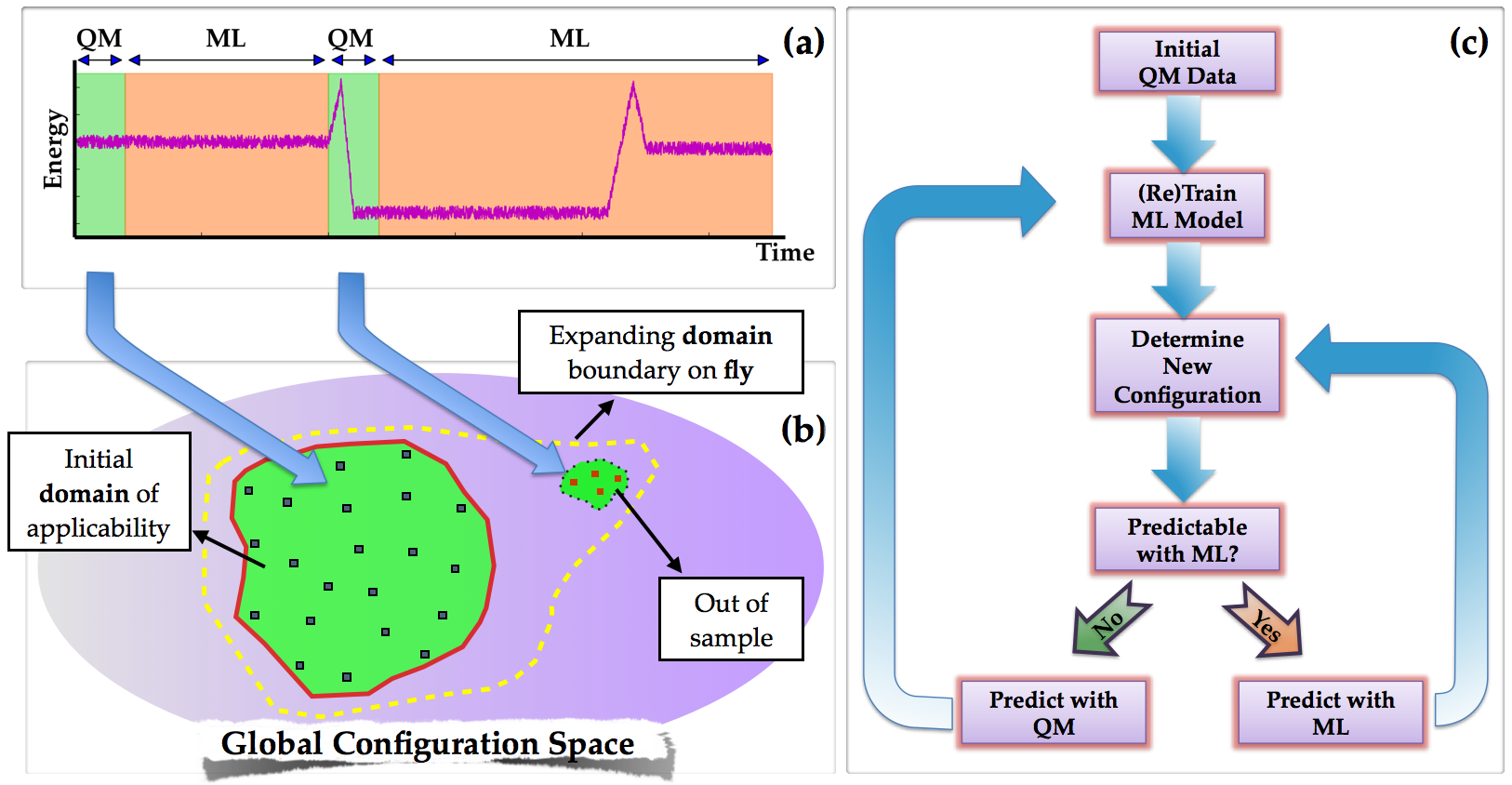}
\caption[Figure1] {(a) A typical MD energy trajectory, with the green and orange regions identifying the quantum mechanical (QM) and machine learning (ML) phases, respectively, of the adaptive learning framework. (b) Expansion of the domain of applicability on-the-fly, if and when new configurations are visited. (c) A flowchart of the adaptive learning framework. The green and orange arrows indicate the use of QM or ML models.  
\label{Figure 1}}
\end{figure*}    

The present contribution provides a pathway for a new solution to the \textit{ab-initio} MD acceleration problem that preserves the fidelity of both QM and the clock. First, we make three observations. 
\begin{enumerate}
\item	During a typical MD trajectory, a system is largely exploring similar configurations, and new features or events are encountered rarely, as schematically portrayed in Figure \ref{Figure 1}(a). This observation is quite universal, and applies to many important processes such as defect diffusion in solids or surface chemical reactions. Taking point defect diffusion as an example, the actual site-to-site hopping of the defect is a \emph{rare event}, while the vibrational motion of the defect (and its surroundings) in its local minimum occupies most of the time and leads to a plethora of “similar” configurations. 
\item	It is fair to assume that similar configurations will have similar properties (such as energies, atomic forces, etc.). If a robust numerical representation of the configurations can be developed, a quantitative measure of (dis)similarity of configurations can be defined, which can then be mapped to (dis)similarities between properties via a learning algorithm. Within the context of accelerated MD, such a machine learning (ML) procedure can be used to predict the energies and forces of similar configurations along the MD trajectory rapidly, provided QM “training” data pertaining to the initial part of the trajectory is available. This is also shown in Figure \ref{Figure 1}(a). 
\item	When a completely new configuration or event is encountered, a decision has to be made to switch back from ML to QM. Most importantly, the new configurations and properties should be included in the learning framework on-the-fly as illustrated in Figure \ref{Figure 1}(b), making the learning process \emph{adaptive}, and continuously evolving with progressive improvement in predictive quality. If this can be accomplished, then, the next time a similar rare event is encountered, QM is unnecessary. This aspect is also captured in Figure \ref{Figure 1}(a).
\end{enumerate}
 
Thus, the basic premise of the proposed strategy is that the significant redundancies implicit in conventional \textit{ab-initio} MD schemes can be systematically eliminated. The flowchart shown in Figure \ref{Figure 1}(c) summarizes the proposed on-the-fly adaptive ML strategy to accelerate \textit{ab-initio} MD. 

It is worth noting that ML strategies are making significant inroads into various aspects of materials science \cite{Rampi_1}, including accelerated and accurate predictions (using past historical data) of phase diagrams \cite{Srinivas_1,Long_1}, crystal structures \cite{Hautier_1,Fischer_1,Zhang_1,Oganov_1}, and material properties \cite{Rupp_1,Pilania_1}, mapping complex materials behavior to a set of process variables \cite{Balachandran_1,Bucholtz_1, Castelli_1}, data analysis of high-throughput experiments \cite{Long_1,Morgan_1,Kusne_1}, etc. Of particular relevance to the present contribution are recent successful efforts that exploit ML methods (neural networks \cite{Behler_2} and Gaussian approximation kernels \cite{Bartok_2}) to develop accurate force-fields (or interatomic potentials) that can allow for significant extension of the time- and length-scales of MD simulations. Nevertheless, the present contribution is one of the first attempts in which the implementation of an adaptive on-the-fly learning scheme to accelerate \textit{ab-initio} MD is discussed. 

The proposed strategy, as captured in Figure \ref{Figure 1}(c), involves a number of vital ingredients. These include: (1) a rigorous and generalizable scheme to represent atomic configurations by continuous numerical fingerprints that are invariant to translations, rotations and permutations of like atom types (as such transformations lead to equivalent configurations); (2) a robust learning algorithm that can map the fingerprints to properties; (3) a decision engine that queries whether the properties of a new configuration are predictable using the current learning model; and (4) needless to say, \textit{ab-initio} (re)training data from the initial part of the MD trajectory and at points when the decision engine makes \textit{ab-initio} calculations mandatory. 

\begin{figure*}
\includegraphics[width=7in]{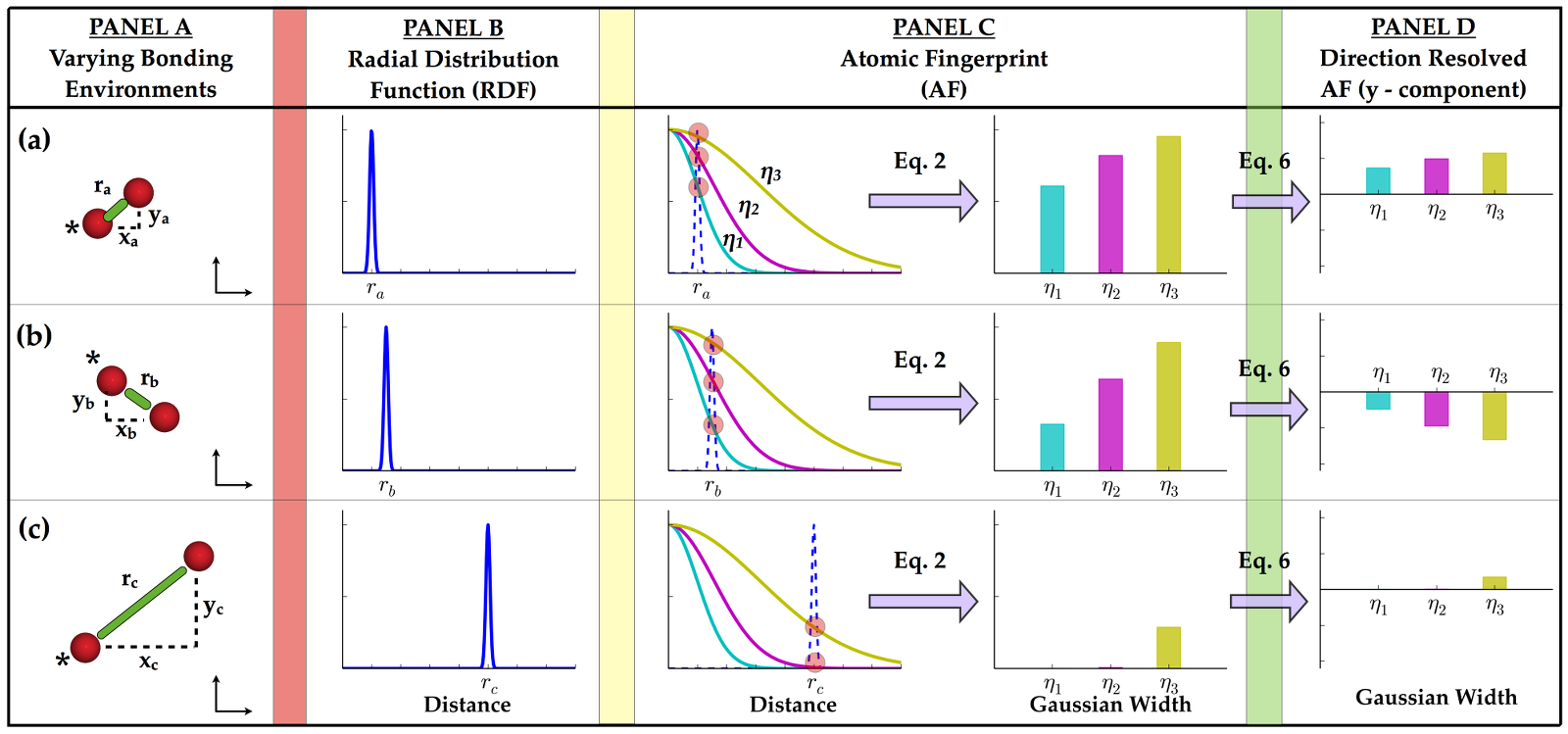}
\caption[Figure2] {Panel A: A homonuclear diatomic molecule displaying three different bond lengths. Panel B: The corresponding radial distribution function (RDF) for each of the bonding environments. Panel C: Transformation of the RDF using Gaussian fucntions on an eta-grid as indicated by the colored lines, into an atomic fingerprint. Panel D: The y-component of the direction resolved atomic fingerprint of an atom in the three bonding environments. The fingerprints generated are for the atom indicated by * in Panel A.
\label{Figure 2}}
\end{figure*}

A firm understanding of the requirements and the limits of the four ingredients listed above is necessary for the practical realization of a high-fidelity, accelerated MD simulation scheme. To directly address this, in this manuscript, we consider fcc Al, a model elemental metallic system in several chemically distinct environments, including (i) defect-free bulk Al, (ii) bulk Al containing a vacancy, (iii) clean (111) Al surface, and (iv) the (111) surface with an Al adatom. For each of the 4 cases above, robust numerical configurational fingerprints are created that allow for high-fidelity predictions of energies and forces at chemical accuracy via a similarity-based learning algorithm. Also, a simple decision engine is presented that detects the occurrence of a new configuration not already in the initial training data set, thus signaling when a fresh QM calculation is required. The combination of the individual working entities should lead us to the ultimate goal of an adaptive learning framework to significantly accelerate \textit{ab-initio} MD simulations on-the-fly. 

\section{\label{sec:level2} Methods and Models}

\subsection{\label{sec:level2a} Fingerprints: Numerical representations of atomic, molecular and crystal environments}

The first critical step in the proposed learning approach is to represent the chemistry and geometry of our system numerically (hopefully, uniquely), such that a mapping can be established between this numerical representation and the property of interest (namely, the energy or forces). Such a representation is referred to here as a “fingerprint” (also commonly referred to as the \textit{feature vector} by the ML community). In what follows, we distinguish between atomic fingerprints and crystal (or molecular) fingerprints. The former captures the coordination environment of a particular atom, while the latter describes the entire ensemble of atoms that are contained within a repeating unit cell (or a molecule). The atomic fingerprint is necessary to predict atomic properties (e.g., forces), while the crystal fingerprint is appropriate to capture global properties (e.g., energy within quantum mechanical schemes, band gap, etc.).  

The atomic or crystal fingerprint is required to satisfy certain requirements \cite{Bartok_1, Yang_1}. In order to adequately capture variations in energy and forces with geometry differences, the fingerprint has to be continuous with respect to slight changes in configuration. Moreover, transformations such as translations, rotations and permutations of atoms of the same type that lead to equivalent systems should not alter the fingerprint.

We first consider atomic fingerprints with the expectation that crystal fingerprints can be built from the constituent atomic fingerprints. A natural first choice for the atomic fingerprint of an elemental system could be the radial distribution function (RDF) defined as follows for a particular atom $i$
\begin{equation}
R_i(r) = \sum\limits_{j\neq i} \delta(r-r_{ij})
\end{equation}

\noindent where $\delta(r)$ is the Dirac delta function and $r_{ij} = |\boldsymbol{r}_i - \boldsymbol{r}_j|$, with $\boldsymbol{r}_i$ being the vectorial position of atom $i$. The sum runs over all the neighboring atoms within an arbitrarily large cutoff distance from atom $i$. Clearly, the RDF, $R_i(r)$, satisfies both the fingerprint requirements mentioned above. The values of $R_i$ in a radial grid can thus be viewed as a numerical fingerprint (or feature vector) describing the coordination environment. Moreover, $R_i(r)$ also captures the geometry in a visually appealing manner. This is demonstrated in Figure \ref{Figure 2}. Panel A contains three homonuclear diatomic molecules (labelled a, b and c) used here to illustrate our fingerprint choices, and Panel B shows the corresponding RDFs. Clearly, the similarity between the bond distances of molecules a and b, and their dissimilarity with that of molecule c is reflected by the corresponding RDFs.  Nevertheless, while these (dis)similarities are apparent to a human, it may not be so for a machine. Typical measures of (dis)similarity utilize the Euclidean norm of the difference between the fingerprint vectors or the dot product between the fingerprint vectors. Clearly, such measures will fail to capture the similarity between molecules a and b, and their dissimilarity with respect to molecule c (as the Euclidean norms of the difference between any pair of the three fingerprint vectors is the same constant value, and the dot products between any pair is zero).

Extending the RDF in a particular way can circumvent the above problem. Rather than using the RDF itself, a transformed quantity can be defined as the integral of the product of $R_i(r)$ and a Gaussian window function

\begin{equation}
G_i(\eta) = \int R_i(r) \ e^{-{\left(\frac{r}{\eta}\right)}^2}\, dr = \sum\limits_{j\neq i} e^{-{\left(\frac{r_{ij}}{\eta}\right)}^2}
\end{equation}
can be used, where $\eta$ is a parameter that describes the extent of the window function. $G_i(\eta)$ is essentially a ``cumulative" version of $R_i(r)$. This is visually demonstrated in Panel C of Figure \ref{Figure 2}, for three $\eta$ values. While $R_i(r)$ is defined in a radial grid, $G_i(\eta)$ is defined in a $\eta$-grid. In order to account for the diminishing importance of atoms far away from the reference atom $i$, we multiply the summand of $G_i(\eta)$ by a cutoff function $f(r_{ij})$ that smoothly vanishes for large $r_{ij}$ values, resulting in our choice of the atomic fingerprint (AF) function, $A_i(\eta)$, given by

\begin{equation}\label{eq:atom_crystal}
A_i(\eta) =  \sum\limits_{j\neq i} e^{-{\left(\frac{r_{ij}}{\eta}\right)}^2} f{(r_{ij})}.
\end{equation}

We note that $A_i(\eta)$ is essentially the radial symmetry function proposed earlier by Behler et al\cite{Behler_1}. Following that previous work we define $f(r_{ij})$ as

\begin{equation}
f{(r_{ij})} =\begin{cases}
   0.5\left[\cos\left(\frac{\pi r_{ij}}{R_c}\right)+1\right] & \text{if } r_{ij} \leq R_c \\
   0       & \text{if } r_{ij} > R_c
\end{cases}
\end{equation}

\noindent where $R_c$ is the cutoff radius, chosen here to be 8 \AA. Interestingly, the $\eta$-grid does not have to be as fine as the radial grid. More importantly, $A_i(\eta)$ does not have the issues that $R_i(r)$ has, with respect to capturing the (dis)similarity between actual physical situations as defined by Euclidean norms. This can be ascertained by inspecting Panel C of Figure \ref{Figure 2}. 

For the molecular or crystal fingerprint (i.e., the fingerprint of the entire molecule or unit cell, $C(\eta)$, also defined on a $\eta$-grid) to be used for mapping the total potential energy of a configuration, we use the average of the atomic fingerprint $A_i(\eta)$ over the constituent atoms, as given by

\begin{equation}\label{eq:crystal}
C(\eta) = \frac{1}{N} \sum\limits_i^N A_i(\eta)
\end{equation}
where N is the total number of atoms in the molecule or unit cell.

Finally, we consider the extension of the $A_i(\eta)$ definition so that it becomes applicable to represent vectorial atomic quantities such as forces. This can be simply done by \emph{resolving} each term in the summation of $A_i(\eta)$ into its Cartesian components, leading to the direction-resolved atomic fingerprints, $\boldsymbol{V}_i(\eta) = \{V_i^x(\eta), \ V_i^y(\eta), \ V_i^z(\eta)\}$ as follows

\begin{equation}\label{eq:atom}
V_i^k(\eta) = \sum\limits_{j \neq i} \frac{r_{ij}^k}{r_{ij}} \ e^{-{\left(\frac{r_{ij}}{\eta}\right)}^2} f{(r_{ij})}, \ k \in \{x,y,z\}  
\end{equation}
where $r_{ij}^k$ is the $k$-th component of $(\boldsymbol{r}_i - \boldsymbol{r}_j)$. Panel D of Figure \ref{Figure 2} visually demonstrates the $V_{i}^y(\eta)$ function for the homonuclear diatomic molecular systems of Panel A. In order to extend the atomic fingerprint (be it $A_i(\eta)$ or $\boldsymbol{V}_i(\eta)$), to non-elemental systems, one could follow a similar approach as above, whereby the atomic fingerprint contains components, one for each atom type.

\subsection{\label{sec:level2b} Learning Method: Kernel ridge regression}

The second critical step is the choice of the learning method. In this work, we have chosen the kernel ridge regression (KRR) technique, which has been used successfully in the recent past within the materials and chemical sciences\cite{Rupp_1,Pilania_1,Kusne_1}. KRR transforms the input fingerprint into a higher dimensional space whereby a linear relation between the transformed fingerprint and the property of interest can be established\cite{Hoffman_1, Muller_1, Hastie_1}. To be precise, the mapping process between the fingerprint and property involves the ``distances" between fingerprints rather than the fingerprints themselves. KRR may thus be viewed as a similarity-based learning method, i.e., similar fingerprints will lead to similar properties.

Within KRR, the property of a system $u$ is given by a sum of weighted Gaussians,
\begin{equation}\label{eq: krr}
P_u = \sum\limits_{v} \alpha_v e^{-\frac{1}{2}{\left(\frac{|d_{uv}|}{\sigma}\right)}^2}
\end{equation}
where $v$ runs over all the cases in the training dataset. $d_{uv}$ is the Euclidean distance between the fingerprint vectors of systems $u$ and $v$. The coefficients $\alpha_v$s and the parameter $\sigma$ are determined during the training phase, whence the objective function $\sum\limits_{v}(P_v - P_v^{QM}) \ + \  \lambda \sum\limits_{v} \alpha_v^2$ is minimized. $P^{QM}_v$ is the QM value of the relevant property, and $\lambda$ is a regularization parameter that should be carefully chosen to avoid overfitting \cite{Rupp_1,Pilania_1}. The parameters $\sigma$ and $\lambda$ are determined by k-fold cross-validation (in this work k=5) on the training dataset. In this method, the training dataset is split into k bins. Each bin acts a new test dataset, whilst the remaining k-1 bins are combined into a new training dataset. The processs is repeated for every bin in the k bins, and for every $\sigma$ and $\lambda$ on a pre-selected logarithmically scaled fine grid. The optimal $\sigma$ and $\lambda$ parameters (i.e., ones that lead to the lowest k-fold cross validation error) are then used in the final model development stage to determine the $\alpha_v$ values for the entire training dataset. At this point, the machinery is in place to predict the property value using Eq. \ref{eq: krr}.

\subsection{\label{sec:level2c} Decision Engine: Fingerprint range}
The third critical step is the decision engine that guides prediction machinery choice (either QM or ML) for energy and force evaluations. If a simulation spends a majority of the time using the \textit{ab-initio} engine, it nullifies any speedup. This raises an important question, namely, how do we judge whether the property of a new configuration can be predicted with the ML approach? One way to classify a new stucture is to compare its fingerprint with those in the training dataset (once sufficient intial training data has been accumulated). If every component of a new fingerprint lies within the range of components of fingerprints already in the training dataset, then we decide that we are in the predictable domain. If not, a fresh QM calculation is mandatory. The new results should then be included in the training set and retraining must be performed to improve the predictive capability. Certainly, a more complex decision engine can be developed by taking inspiration from the field of domain applicability as used within drug prediction\cite{Kaneko_1, Carrio_1, Sheridan_1}, but this is not attempted here.

\subsection{\label{sec:level2d} Data Generation: Quantum mechanics}
 
Data for the 4 cases (i) defect-free bulk Al, (ii) bulk Al containing a vacancy, (iii) a clean (111) Al surface, and (iv) the (111) surface with an Al adatom was generated from \textit{ab-initio} (DFT) MD runs in a micro-canonical ensemble (NVE) using a timestep of 0.5 fs, with the Vienna \textit{ab-initio} Simulation Package\cite{Kresse1,Kresse2}. The bulk cases (i and ii) consisted of a 32 (or 31 with the vacancy) atom model. The surface cases (iii and iv) consisted of a 16 (or 17 with adatom) atom surface model. The generalized gradient approximation (GGA) functional parametrized by Perdew, Burke and Ernzerhof (PBE) to treat the electronic exchange-correlation interaction, the projector augmented wave (PAW) potentials, and plane-wave basis functions up to a kinetic energy cutoff of 520 eV were employed\cite{Blochl1,Perdew1,Kresse1,Kresse2}. A $\Gamma$-centered k-point mesh of 7x7x7 and 7x7x1 were used for the bulk and surface calculations, respectively. 

\subsection{\label{sec:level2e} Training and Test Datasets}

As a reminder, we note that two types of fingerprints are used in this work: $C(\eta)$ to map to total potential energies and $\boldsymbol{V}_i(\eta)$ to map to atomic forces. Using \textit{ab-initio} MD, a total of 2000 configurations were generated for each of the 4 material systems considered. For the energy prediction assessment, various amounts of training data was randomly selected from the above sampled configurations, while the remaining was considered as the test dataset, used to gauge model performance. Similarly for the force prediction assessment, various amounts of training data was randomly selected from all the atomic environments sampled (i.e., 64000 for case i and ii, and 32000 for case iii and iv), with the remaining considered as the test dataset.

\section{\label{sec:level3} Desired Parameter Choices}

The specific choice of model parameters is critical to performance\cite{Hansen_1}. To establish fidelity in predictions, we extensively tested two key quantities: (1) the length of the fingerprint vector (i.e., number of points in the $\eta$ grid), and (2) the training dataset size.

\subsection{\label{sec:level3a} Fingerprint Vector Size}

\begin{figure}
\includegraphics[width=2.5in]{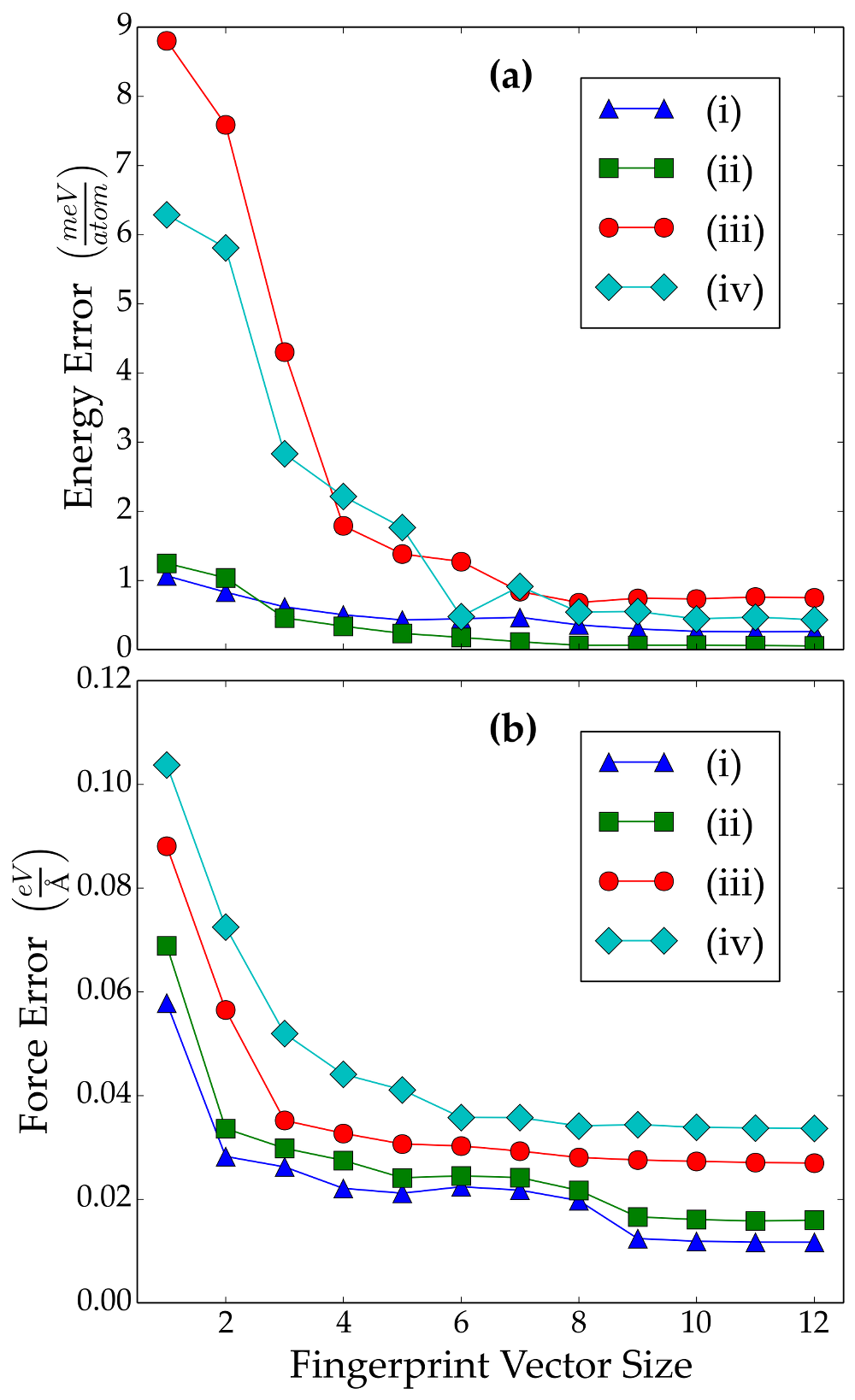}
\caption[Figure3] {Energy (a) and force (b) error versus length of fingerprint size for (i) defect-free bulk Al, (ii) bulk Al containing a vacancy, (iii) a clean (111) Al surface, and (iv) the (111) surface with an Al adatom.
\label{Figure 3}}
\end{figure}

A natural question that arises is, how dense should the $\eta$-grid defined in Eq. \ref{eq:atom_crystal} and \ref{eq:atom} be, to adequately describe the varying atomic and crystal environments encountered. To critically address this question, we systematically increased the number of $\eta$ values from 1 to 12 (thus increasing fingerprint complexity). Starting with small $\eta$ values, as this captures the dominant nearest neighbor shell contributions, we added more components to the fingerprint based on a logarithmic $\eta$ grid between 10$^{-1}$ \AA \ and 10$^{2}$ \AA. For each case, we used a training dataset size of 100 and 500 for the energy and force models, respectively (these sizes are shown in the next subsection to be sufficient to ensure convergence of the predictions). The model error as shown in Figure \ref{Figure 3}, decreases with increasing fingerprint complexity for all 4 cases, suggesting that convergence has been achieved. 

Interestingly, with the energy model the fingerprint complexity is also dependent on the type of structure being studied. As seen in Figure \ref{Figure 3}(a), in order to achieve chemical accuracy in energy (MAE $<$ 1 {\small{$\rm{\frac{meV}{atom}}$}}), the bulk cases (i and ii) required a 3-component fingerprint whereas the surface cases (iii and iv) required an 8-component fingerprint. The above observation is not entirely surprising. A surface model, unlike the bulk, is non-periodic along the surface normal whereby atoms of varying coordination exist, depending on the atom position (surface or below). The learning algorithm maps the energy to a crystal fingerprint (which is averaged across all atoms), and hence the resolution of each individual atom is smeared out. Only upon increasing fingerprint complexity can we achieve an accurate model. Such a concern does not exist for the force model, since a one-to-one mapping between the atomic environment and the force is undertaken. It is for this reason that the force error, as seen in Figure \ref{Figure 3}(b), for all 4 cases starts high (MAE $>$ 0.05 {\small{$\rm{\frac{eV}{\AA}}$}}) and decreases systematically, with error levels converging well below numerical DFT noise.  

\subsection{\label{sec:level3b} Training DataSet Size}

\begin{figure}
\includegraphics[width=2.5in]{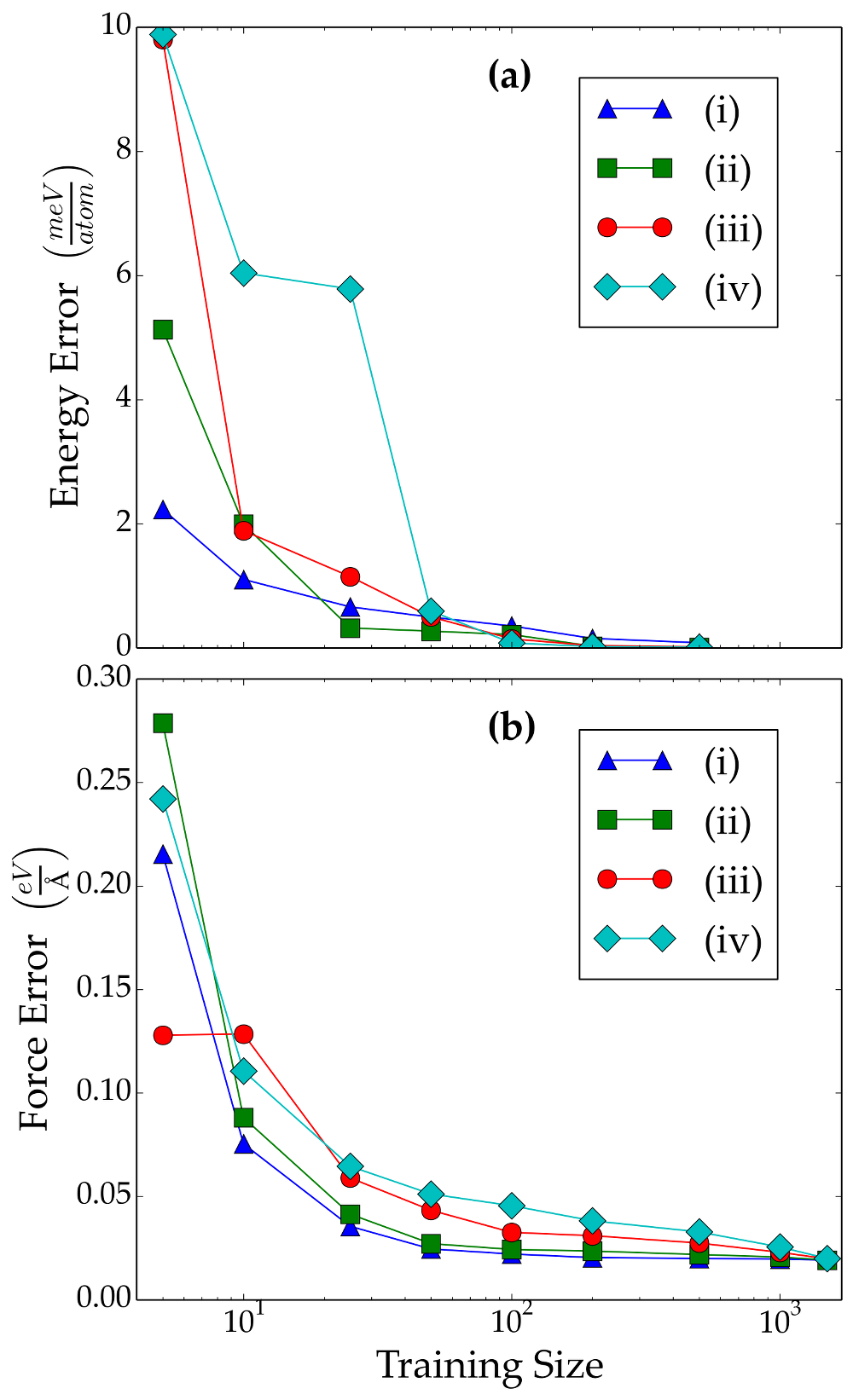}
\caption[Figure4] {Energy (a) and force (b) error versus training size for (i) defect-free bulk Al, (ii) bulk Al containing a vacancy, (iii) a clean (111) Al surface, and (iv) the (111) surface with an Al adatom.
\label{Figure 4}}
\end{figure}

Another aspect critical to the performance of the learning algorithm, is the size and choice of training data. Kernel ridge regression, unlike traditional neural network, is not prone to overfitting when proper cross-validation measures are undertaken. More the data the better the model, although practically a finite dataset is used, as computational overhead relates to the $\mathcal{O}(n^3)$ with training dataset size\cite{Witten_1}. To determine the optimal training size that balances computational expense with accuracy, model error versus training dataset size was studied as shown in Figure \ref{Figure 4}, using an 8-component crystal and direction-resolved atomic fingerprint. Clearly, a systematic decrease in error with increased training once again signifies convergence. 

Models with small training dataset sizes ($<$ 25 for energy and $<$ 50 for force) leads to poor learning, resulting in high errors. For the energy model, bulk cases (i and ii) require 25 configurations or more, while the surface cases (iii and iv) require 50 configurations or more to achieve error convergence. On the other hand for the force model, the bulk cases converge to the desired accuracy with $<$ 50 training configurations, while the surface cases require $>$ 200 configurations. Similar to the observations with fingerprint complexity, as the configurational expanse increases from the bulk to surface owing to the non-periodicity, the training size required increases accordingly.

\section{\label{sec:level4} Prediction of Energy and Forces}

\begin{figure*}
\includegraphics[width=7in]{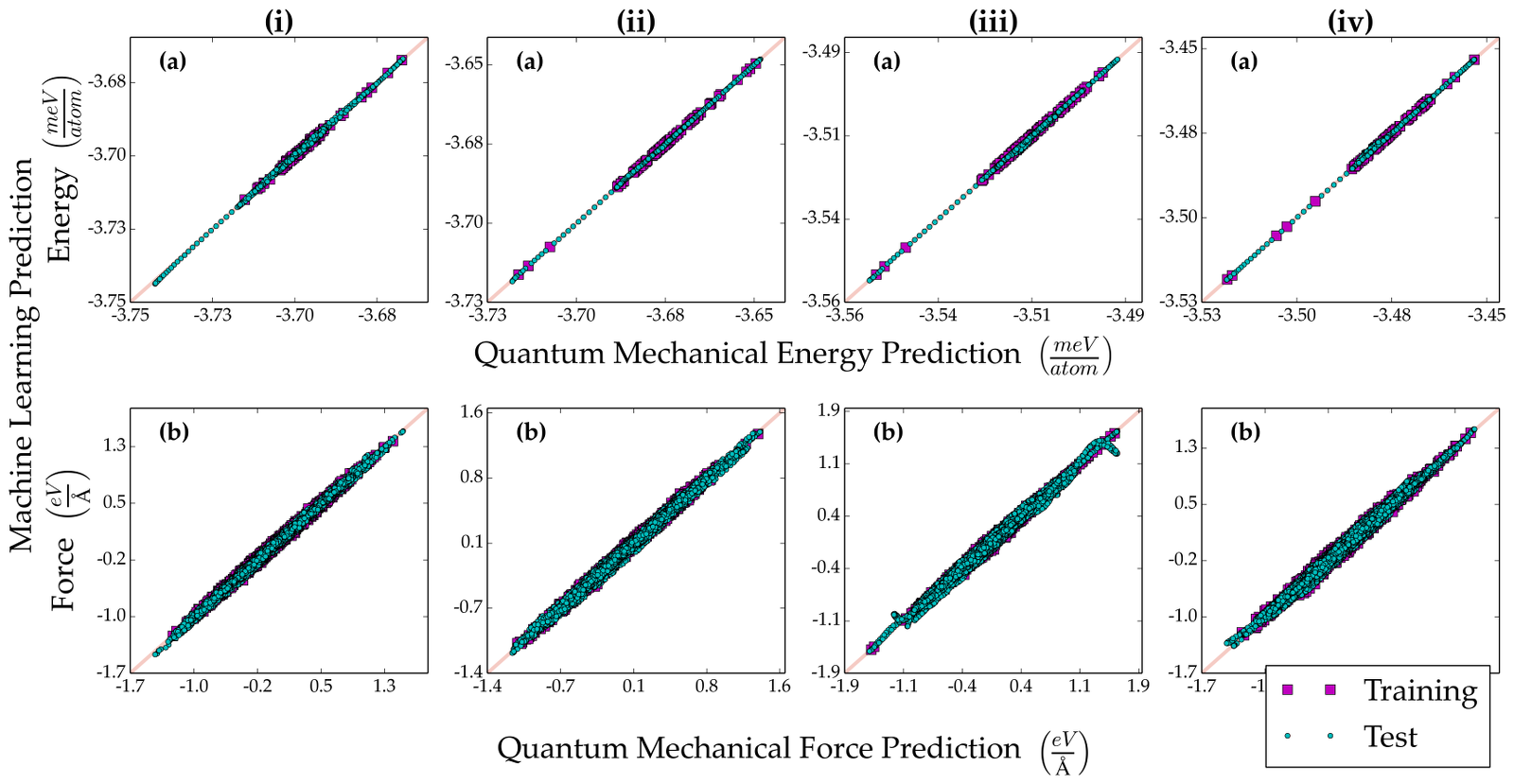}
\caption[Figure5] {Parity plot for (i) defect-free bulk Al, (ii) bulk Al containing a vacancy, (iii) a clean (111) Al surface, and (iv) the (111) surface with an Al adatom, with energy (a) and force (b) predictions in the top and bottom rows, respectively. An 8 component fingerprint, with 100 training configurations for the energy models and 100 (for (i) and (ii)) and 750 (for (iii) and (iv)) training configurations for the force models were used. 
\label{Figure 5}}
\end{figure*}

Based on the convergence studies of model parameters in Sec. \ref{sec:level3}, we chose 8 components for both the crystal fingerprint ($C(\eta)$) and direction-resolved atomic fingerprint ($\boldsymbol{V}_i(\eta)$). Secondly, 100 training configurations for energy and 100 (for (i) and (ii)) or 750 (for (iii) and (iv)) training configurations for the force model, were randomly selected. Using the above parameters as input to the learning algorithm, we predict energy and forces for the 4 test cases of elemental Al, as shown in Figure \ref{Figure 5}. Each prediction takes a fraction of a millisecond. Our ML predictions agree well with the QM data, with the observed errors ($<$ 1 {\small{$\rm{\frac{meV}{atom}}$}} and $<$ 0.05 {\small{$\rm{\frac{eV}{\AA}}$}}) reported in Table \ref{Table1}. This suggests well learned models in all the cases. Errors of this magnitude are comparable to errors arising within the approximations made within DFT itself. It is accuracy at this level that allows us to bypass expensive QM methods and rely upon the proposed learning approach for quick energy and force predictions. However, to build a self evolving learning method (as we propose in Figure \ref{Figure 1}(c)), that adapts during the course of a simulation requires a scheme able to recognize situations that are outside the original training domain. 

\begin{center}
\begin{table}
\caption{Mean absolute error in energy and force predictions of the 4 cases. Test error in bold and training error in brackets.}
\label{Table1}
\bgroup
\def\arraystretch{1.5}
\begin{tabular}{lccccccc}
\hline\hline
\multicolumn{1}{l}{Case} & \multicolumn{3}{c}{Energy \small{($\rm{\frac{meV}{atom}}$)}} & \multicolumn{4}{c}{Force \small{($\rm{\frac{eV}{\AA}}$)}} \\
\hline

(i) Defect-free bulk Al	     && \textbf{0.04} (0.03) && \textbf{0.02} (0.02) \\
(ii) Bulk Al w. vacancy	     && \textbf{0.06} (0.02)	&& \textbf{0.02} (0.02) \\
(iii) Clean (111) Al surface	 && \textbf{0.16} (0.08)	&& \textbf{0.03} (0.02) \\
(iv) (111) Surface w. adatom	 && \textbf{0.22} (0.07)	&& \textbf{0.03}	(0.03) \\
\hline
\end{tabular}
\egroup
\end{table}
\end{center}

\section{\label{sec:level5} Decisions on predictability}

\begin{figure}[b]
\includegraphics[width=3.5in]{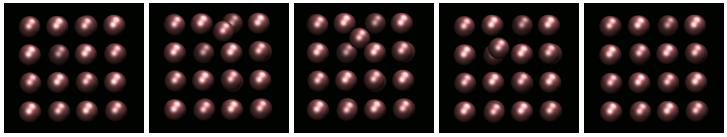}
\caption[Figure6] {Vacancy migration within bulk Al. The structures shown correspond to s.png 1, 5, 10, 15, 20 along the 20-step trajectory. 
\label{Figure 6}}
\end{figure}

\begin{figure*}
\includegraphics[width=6in]{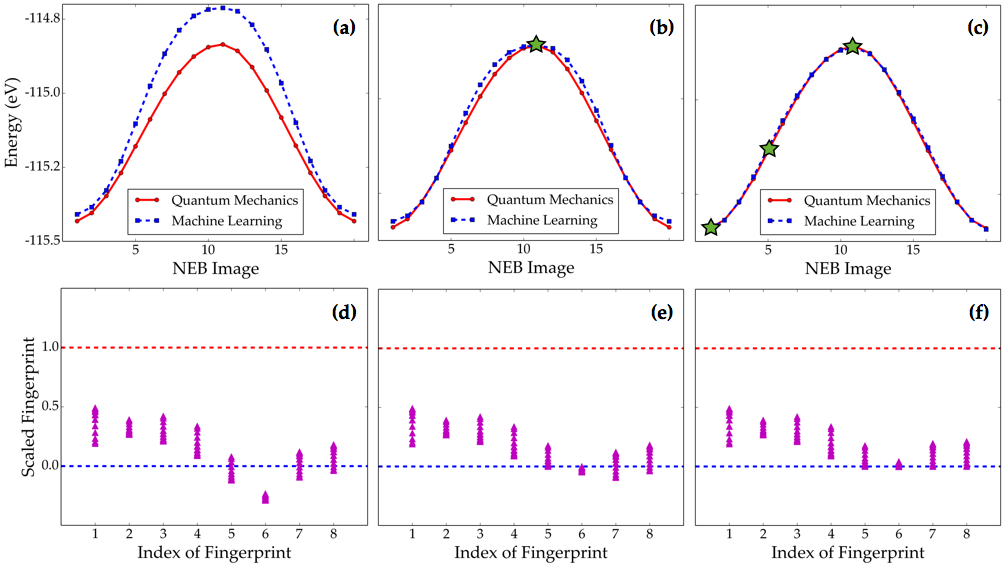}
\caption[Figure7] {QM and ML energy, (a)-(c), and the range of crystal fingerprint components with respect to the training data set, (d)-(f), of each image along the vacancy migration trajectory. (a) and (d) with no retraining, (b) and (e) with the TS added to training and (c) and (f) with TS and image 1 and 5 added to the training. $\star$ indicates the configurations added during retraining.
\label{Figure 7}}
\end{figure*}

ML methods are, in general, interpolative and are unable to handle situations outside the training domain. To demonstrate such a situtation within the context of this work, a series of configurations that mimic the migration trajectory of a vacancy in bulk Al were generated as shown in Figure \ref{Figure 6}. The energy and forces for each configuration along the migration trajectory were predicted using QM and our ML model. Given the short time span explored while generating the training data (case ii), no such migration event was actually observed. Thus, configurations close the transitions state (TS) should be inaccurately predicted by ML. 

\begin{figure}
\includegraphics[width=2.19in]{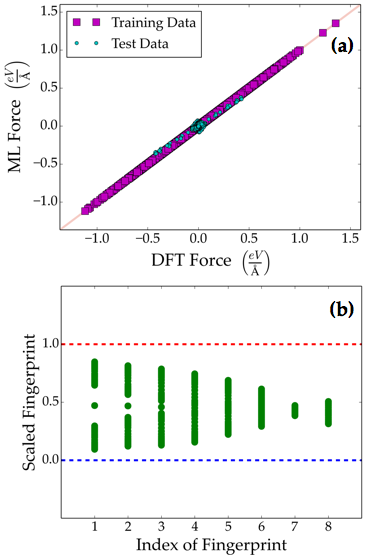}
\caption[Figure8] {(a) Direction resolved atomic fingerprint range compared to the training dataset of the force model, and (b) Parity plot showing accurate force prediction without any retraining. 
\label{Figure 8}}
\end{figure}

Figure \ref{Figure 7}(a) plots the true (QM) and predicted (ML) energy of each configuration along the migration trajectory, with the TS at the apex. Clearly, the starting and ending configurations are predicted well (as they resemble those in the training dataset). However, the error increases significantly as we move towards the TS, as these configurations were never sampled during training. Upon adding just the TS configuration to the training database and retraining, the error along the entire trajectory drops within acceptable accuracy (Figure \ref{Figure 7}(b)). Adding more configurations along the migration pathway to the training dataset and retraining further refines the energy predictions even more (Figure \ref{Figure 7}(c)). The configurations added for retraining are indicated by $\star$ in Figure \ref{Figure 7}(b) and \ref{Figure 7}(c). Interestingly, as can be seen in Figure \ref{Figure 8}(a), the atomic forces of all configurations along the trajectory are accurately predicted with error $<$ 0.05 {\small{$\rm{\frac{meV}{atom}}$}}, \textit{without any retraining}.

To illustrate how one can detect whether the properties of a stucture are predictable or not, we used the method discussed in Sec. \ref{sec:level2c}. A plot of the relative location of each crystal fingerprint component compared with the training dataset bounds (maximum and minimum value given by the red and blue dotted lines is shown in Figure \ref{Figure 7}(d)). In the retrained models (only including the TS configuration, Figure \ref{Figure 7}(e), and including the TS with other configurations, Figure \ref{Figure 7}(f)) the crystal fingerprint components approach the training dataset bounds, and the error drops as a result. With the forces however, all the atomic fingerprint components in the migration trajectory fall within training dataset bounds even before training, as shown in Figure \ref{Figure 8}(b). Therefore, the predicted force errors are negligible as seen in the parity plot of Figure \ref{Figure 8}(a). The proposed decision engine is a rudimentary but an effective approach to recognize structures which may fall outside the original training domain. 

\section{\label{sec:level6} Implications of this work}

Thus far, we have demonstrated that energies and atomic forces may be predicted with chemical accuracy using a ML algorithm trained on QM data. Critical to this capability is the representation of atomic configurations and environments using continuous numerical fingerprints. Here, we have presented a class of simple, intuitive, efficient and elegant fingerprints that can capture scalar (e.g., energy) and vector (e.g., force) quantities. We also presented a scheme that can recognize new cases not already in the training domain, which can subsequently be included in the training process thus making the prediction scheme adaptive and the predictive power monotonically increasing in quality.

All the ingredients required to eliminate (expensive) redundancies that plague \textit{ab-initio} MD simulations and hence accelerate them significantly are thus in place. The scheme proposed here, shown in Figure \ref{Figure 1}(c), closely integrates with an existing DFT code; this will allow the learning scheme to become adaptive on-the-fly, and significantly mitigate the time-scale challenge that \textit{ab-initio} MD schemes currently face. As several such simulations are performed for a particular system, the accumulated information (i.e., fingerprints, forces and energies), if diverse, can lead to the creation of a force-field, using which subsequent simulations can be performed without the need for an explicit DFT engine (this is in the spirit of recent ML based force-field development efforts\cite{Behler_2,Bartok_1}). Indeed, this is particularly true with the forces and the force fingerprints, $\boldsymbol{V}_i(\eta)$, which are purely functions of the atomic environment, unlike the total potential energy and the crystal fingerprint, which are functions of the supercell as a whole. Thus, a scheme purely based on the forces (which is conceivable as energies can be obtained from the forces through integration) does not have to be linked to a particular supercell. Such a development can mitigate the length-scale challenge faced by \textit{ab-initio} MD.

The present work may also impact non-MD simulations. For instance, structure prediction schemes require either total potential energies or total potential energies and forces \cite{Oganov_1,Goedecker_1,Wales_1}. A scheme analogous to the flowchart of Figure \ref{Figure 1}(c) can be conceived for an adaptive on-the-fly ML scheme to accelerate structure prediction calculations (or even stand-alone schemes once sufficient history is accumulated, as discussed above). Going further, the same paradigm can be applied to map the fingerprints to other local and global properties of interest, such as effective charges, dipoles, polarization, band gap, dielectric constant, etc. Finally, we note that, although the QM training data discussed here came from one flavor of DFT calculations, the present scheme is applicable to any class of data, including beyond-DFT and other more sophisticated QM methods, thus improving of the predictive power further at no extra cost (other than that incurred at the training phase). The implications of the present development are expected to be far reaching.

\section{\label{sec:level7} Summary}

A detailed understanding of the dynamical evolution of materials and processes involves timescales that are beyond the reaches of present day quantum mechanical or \textit{ab-initio} MD methods. The primary causes of the bottlenecks in such approaches are the expensive and repetitive energy and force computations required, and the small times.png involved. Acceleration schemes proffered thus far either do not preserve the fidelity of the time evolution, or have very limited domains of applicability. 

In this contribution, we presented a scheme that can enormously accelerate MD simulations while still preserving the fidelity of the time-evolution, and allow us to span timescales previously inaccessible at the \textit{ab-initio} level of accuracy. The basic premise of this work is that similar configurations are constantly visited during the course of an MD simulation, and that the redundancies implicit in conventional \textit{ab-initio} MD schemes can be systematically eliminated. The foundations for such an accelerated \textit{ab-initio} MD scheme is laid out here. A machine learning scheme is proposed which learns from previously visited configurations in a continuous and adaptive manner on-the-fly, and predicts the energies and forces of a new configuration at a minuscule fraction of the time taken by conventional \textit{ab-initio} methods. Key elements of this new accelerated \textit{ab-initio} MD paradigm include representations of atomic configurations by numerical fingerprints, the learning algorithm, a decision engine that guides the choice of the prediction scheme, and, of course, the requisite amount of \textit{ab-initio} (re)training data. 

The performance of each aspect of the proposed \textit{ab-initio} MD acceleration scheme is critically evaluated for Al, a model elemental system, in several different chemical environments, including defect-free bulk, bulk with a vacancy, clean (111) surface, and the (111) surface with an adatom. The robust configurational fingerprints utilized, and the learning algorithm adopted lead to energy and force predictions at chemical accuracy, provided sufficient fingerprint components and \textit{ab-initio} training data are used. The simple and intuitive decision engine that guides whether machine learning or quantum mechanics needs to be used to predict the energies and forces of a new configuration is also shown to be robust. When quantum mechanics is mandated, the new results are to be used in a machine learning retraining step; this makes the scheme adaptive on-the-fly. With the above critical pieces in place, we have a complete prescription for a new accelerated \textit{ab-initio} MD paradigm.

The ideas contained within this manucript, although demonstrated for just an elemental metallic system, is readily extendable and applicable to non-metallic as well as non-elemental systems. Even though the focus of the present work is to accelerate \textit{ab-initio} MD simulations, the same adaptive strategy can be applied for the learning and prediction of other properties as well.

\section*{Acknowledgement}
This work was supported financially by a grant from the Office of Naval Research (N00014-14-1-0098). Partial computational support through a Extreme Science and Engineering Discovery Environment (XSEDE) allocation is also gratefully acknowledged. The authors would like to acknowledge useful discussions with Kenny Lipkowitz and Avinash M. Dongare, and a critical reading of the manuscript by Ghanshyam Pilania, Tran D. Huan and Arun Mannodi-Kanakkithodi.


%

\end{document}